\documentclass[aps,prl,twocolumn,groupedaddress,showpacs]{revtex4}

\usepackage{graphicx}
\begin{document}

\title{Asymmetric diffusion in the delta-kicked rotor with broken symmetries}

\author{P. H. Jones, M. Goonasekera, H. E. Saunders-Singer, T. S. Monteiro \& D. R. Meacher}
\email[]{philip.jones@ucl.ac.uk}
\homepage[]{http://lasercooling.phys.ucl.ac.uk}

\affiliation{Department of Physics and Astronomy, University
College London, Gower Street, London, United Kingdom, WC1E 6BT}

\date{\today}

\begin{abstract}
We report an experimental investigation of momentum diffusion in the
$\delta$-function kicked rotor where time symmetry is broken by a
two-period kicking cycle and spatial symmetry by an alternating linear potential. The momentum diffusion constant is thus
modified by kick-to-kick correlations which show a momentum
dependence. We exploit this, and a technique involving a moving
optical potential, to create an asymmetry in the momentum
diffusion that is due entirely to the chaotic
dynamics.
\end{abstract}


\maketitle

Classical systems with chaotic dynamics can exhibit very different behavior in the quantum limit.  One such system that has been frequently studied is the delta-kicked rotor (DKR), modeled using laser-cooled atoms in a pulsed periodic optical potential.  Purely quantum mechanical phenomena such as dynamical localization \cite{Barucha1999} (the quantum suppression of classical momentum diffusion), quantum resonances \cite{Oskay2000} (ballistic rather than diffusive energy growth) have been observed.  Asymmetry in the DKR has been produced via accelerator modes \cite{Oberthaler1999} where the maximum momentum is transferred in each pulse to coherently impart a large amount of momentum to a particle.

Recently DKR systems where the timing of the kicks is subject to change, either by timing noise \cite{Oskay2003} or by using a two-period cycle of kicks with small \cite{Jonckheere2003} or large \cite{Jones2004b} deviation from period one have been studied.  The two cases are interesting to compare in that timing noise was found to destroy classical correlations that give rise to fluctuations in the diffusion rate, whereas for the two-period cycle the classical correlations acquire a momentum dependence.  For the large deviations from period 1 (the 2$\delta$ kicked rotor) new families of correlations were shown to exist and verified experimentally.  

In this paper we report an experimental realisation of the kicked rotor with broken time and space symmetry that was described theoretically in \cite{Jonckheere2003}.  We used cold atoms in a pulsed optical lattice to model the delta kicked rotor with a two-period kicking cycle to break time symmetry.  We use an optical lattice that is moving with constant velocity in the laboratory frame in order to probe the momentum dependence of the momentum diffusion rate in this case.  We then use an accelerating optical lattice to additionally break spatial symmetry. This proof-of-principle experiment is, we believe, the first demonstration of an asymmetric momentum diffusion that is the result of purely chaotic dynamics and does not rely on specific structures in phase space.  

An optical lattice formed by two counter-propagating laser beams
may be used to trap laser-cooled atoms in a one-dimensional
periodic potential \cite{Meacher1998, Grynberg2001}, the
Hamiltonian for which is
\begin{equation}
H = \frac{p^2}{2M} + V_0\cos(2k_L x)
\label{Hamiltonian}
\end{equation}
where $M$ is the mass of the atom, $k_L = 2\pi / \lambda$ the
laser wavevector and $V_0$ the potential depth. If the optical
lattice is applied as a series of short ($\delta$-function) pulses
with period $T$, then we may compare this with the Hamiltonian for
the $\delta$-kicked rotor as written in the usual dimensionless
form:
\begin{equation}
\mathcal{H} = \frac{\rho^2}{2} + K\cos(\phi)\sum\delta(\tau - n)
\end{equation}
where $K$ is the stochasticity parameter which describes the
strength of the kick. Here $\rho = 2Tk_Lp/M$ is a scaled
momentum, $\phi = 2k_Lx$ a scaled position, $\tau = t/T$ a scaled
time and $\mathcal{H} = 8\omega_RT^2H/\hbar$ the scaled
Hamiltonian. The commutation relation $[\phi,\rho] = i8\omega_RT$
gives the scaled unit of system action or effective Planck
constant $\hbar_{ef\!f} = 8\omega_RT$ ($\omega_R$ the recoil
frequency) which may be controlled  through the period of the
pulses. The dynamics of the kicked rotor have been well studied,
particularly through the use of cold atoms in pulsed optical
lattices \cite{Raizen1999}. One important feature is the time-dependence of the
momentum which grows diffusively up to an $\hbar_{ef\!f}$ dependent
time, $t^* \propto \hbar_{ef\!f}^{-2}$, the quantum break time,
before saturating. At times $t < t^*$ to lowest order the
diffusion constant, $D \propto K^2$. Corrections to this arise
from correlations between kicks which appear as Bessel functions \cite{Rechester1980}:
\begin{equation}
D(K) = K^2\bigg[\frac{1}{2} - J_2(K) - J_1^2(K) + J_2^2(K) +
...\bigg]
\end{equation}
the effects of which have been observed as anomalous momentum
diffusion for particular values of $K$ \cite{Klappauf1998}.

In \cite{Jonckheere2003} it was shown that for the $\delta$-kicked
rotor where time symmetry is broken by a two-period kicking cycle
of periods $T(1+b):T(1-b)$ where $b<1$, then for short times these
correlations give rise to a momentum dependent diffusion constant
$D(K, \rho, b) \simeq D_0 - C(2,\rho)$, where $D_0 \simeq
K^2[1/2-J_1(K)^2]$ and $C(2,\rho) = K^2J_2(K)\cos(2\rho b)$
arising from correlations between kick number $i$ and kick number $i+2$. This
correction has a finite lifetime, denoted the ratchet
time, $t_{rat}$ in \cite{Jonckheere2003}, which depends on the parameter $b$ as
$t_{rat}\sim 1/Db^2$. This timescale is different from, and may be
controlled independently of, the break time, which lead to the
main conclusion of \cite{Monteiro2002} that the clearest experimental signature of this phenomenon would require $t^*/t_{rat} \sim Db/\hbar_{ef\!f} \sim 1$.

By including a linear term of alternating sign in the kicking
potential Jonckheere \textit{et al.} also showed that the
$C(2,\rho)$ term may be made locally asymmetric around $\rho$=0,
as $C(2,\rho)\rightarrow K^2J_2(K)\cos(2\rho b -A)$ where $A$ is
the (scaled) gradient of the linear ``rocking" term, even for
parameters where, unlike \cite{Cheon2002} there are no significant
stable structures remaining. It was suggested that this system may
be used to observe an asymmetry in the momentum diffusion or to produce a
chaotic momentum filtering.  An analysis of the Floquet states of this perturbed period kicked rotor may be found in \cite{Hur2004}.

\begin{figure}
\includegraphics[width=0.6\columnwidth]{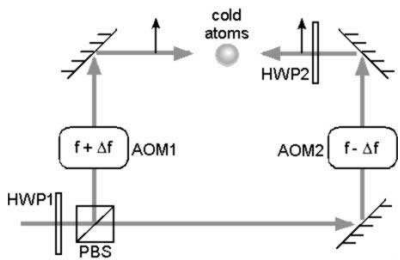}
\caption{Diagram of apparatus\label{ApparatusDiagram}. The
half-wave plate HWP1 and polarizing beam splitter PBS are used to
create two equal intensity beams which are shifted in frequency by
$f\pm \Delta f$ by the AOMs. HWP2 is used to make the
polarizations of the two beams parallel.}
\end{figure}

In our experiment we realize a model of the $\delta$-kicked rotor
using laser-cooled cesium atoms in a far-off resonant pulsed
optical lattice. The lattice is formed by two horizontal
counter-propagating laser beams, $1/e$ radius (0.95$\pm$0.05~{\rm mm}),
with parallel linear polarizations (see
figure~\ref{ApparatusDiagram}) which produces a spatial variation
of the AC Stark shift that is proportional to the local intensity,
and hence sinusoidal.

The pulses are produced by rapidly switching the drive voltage to
the acousto-optic modulators (AOMs) according to a pre-defined
sequence. The time between the kicks may be altered in order to
produce the two-period ``chirped" kicking cycle described above.
The experiment proceeds as follows. Cesium atoms are trapped and
cooled in a standard six-beam magneto-optic trap (MOT) before
further cooling in an optical molasses to an rms scaled momentum
width of $\sigma_{\rho} \simeq 4$. The molasses light is turned
off using an AOM and the periodic ``kicking" potential applied.
The beams for the kicking potential are derived from a Ti:Sapphire
laser with an output power of 1~W at 852~nm, detuned typically 2000
linewidths (natural linewidth $\Gamma$ = 2$\pi \times$ 5.22~MHz) to
the low frequency side of the D2 cooling transition in cesium, which is
sufficient for effects due to spontaneous emission to be
neglected. This is split into two equal intensity beams using a
half-wave plate and polarizing beam splitter (HWP1 and PBS in
figure~\ref{ApparatusDiagram}) and each beam sent through an AOM.
The two AOMs are driven by separate (phase-locked) radio-frequency
synthesizers that are controlled by separate fast
radio-frequency switches but triggered by the same arbitrary
function generator that produces the kicks. After the kicking the
cloud of cold atoms is allowed to expand ballistically for up to 20~ms before a pair of counter-propagating near-resonant
laser beams are switched on and the fluorescence from the atoms
imaged on a CCD camera. From the spatial distribution of the
fluorescence it is then possible to extract the momentum
distribution. Using this apparatus we have checked that the
dynamical localization which is characteristic of quantum chaos
can be observed (for regularly spaced kicks, i.e. $b=0$ in the above) as a growth in momentum for a finite number of
kicks and a change to an exponential momentum distribution.

The momentum dependence of the diffusion constant may be probed by
using a sample of cold atoms with a non-zero mean initial momentum, such
as may be prepared by cooling in an optical molasses in the
presence of a non-zero magnetic field \cite{Shang1990}. A disadvantage of this
technique is that the wings of the atomic distribution may easily
extend beyond the field of view of the CCD camera for relatively
low momentum. Instead we have used a moving optical lattice formed
by laser beams with a controlled frequency difference to make the
kicking potential, so that atoms which are stationary in the
laboratory frame (remain in the center of the CCD picture) have a
non-zero momentum in the rest frame of the optical potential. This
is achieved by driving the AOMs at frequencies $f\pm \Delta f$ as shown in figure~\ref{ApparatusDiagram}, such that the atomic momentum in
the rest frame of lattice is $\rho_L = m\lambda^2 \Delta f
\hbar_{ef\!f}/4\pi \hbar$. Using this technique the mean momentum in
the lattice frame, $\rho_L$, may be varied over a
large range 
in order to sample several periods of
the oscillation of the diffusion constant without the beams
becoming significantly misaligned from the cloud of cold atoms.

We have investigated several values of the parameters $K$ and
$\hbar_{ef\!f}$, but present here those from conditions similar to
those in \cite{Jonckheere2003}, that is $K=3.3$ (10\% error arising manily from the measurement of the beam intensity) and
$\hbar_{ef\!f}=1$ for values of $b=$1/16 and 1/32.  This value of $K$ was chosen as it corresponnds to the first maximum of the Bessel function $J_2(K)$, and hence may be expected to produce the clearest experimental signature, i.e. the largest amplitude oscillations in momentum asymmetry.  Although for the Standard Map phase space is not completely chaotic for $K = 3.3$, the introduction of the parameter $b$ ensures no stable structures remain. 

\begin{figure}
\includegraphics[width=1.0\columnwidth]{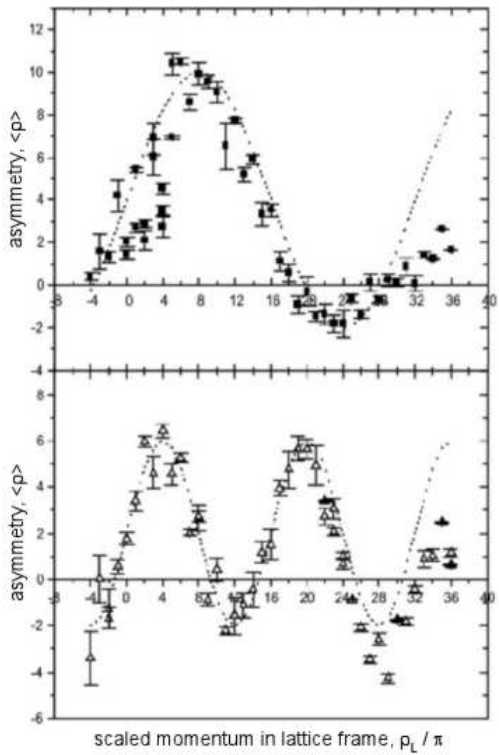}
\caption{Momentum asymmetry \textit{vs} starting momentum in the
lattice frame for $K=3.3$, $\hbar_{eff}=1$, $b=1/32$ (filled
squares) and $b=1/16$ (open triangles).  The dotted lines are sinusoidal with period $\pi /b$ and are intended as a guide only.\label{RatchetResults}}
\end{figure}

For these experiments the period of the kicks is $T = 9.47~\mu{\rm s}$
and pulses are square with duration typically $t_p = 296~{\rm ns}$
($t_p/T = 1/32\leq b$), which is sufficient for there to be no substantial effects on the
diffusion constant due to the finite temporal width of the kicks  in the region of $\rho_L = 0$ \cite{Klappauf1999} (for larger $\rho_L$ these effects become important and start to affect the data). An investigation of the effects
on the asymmetry of momentum diffusion arising the finite width of the kicks was presented in \cite{Jones2004c}.  

We characterise the asymmetry of the momentum distribution after the kicks by the first moment of the distribution, $\langle \rho \rangle = \int \rho N(\rho)d\rho/\int N(\rho)d\rho$, and plot this as a function of $\rho_L$ as shown in figure~\ref{RatchetResults}.  We observe that the asymmetry for each experiment oscillates with a period $\pi/b$ in agreement with the theory of~\cite{Jonckheere2003}, and shown by the dashed lines which are $\propto \sin(32\pi \rho_L)$ (top panel of figure~\ref{RatchetResults}) and $\propto \sin(16\pi \rho_L)$ (lower panel).  The data appears to deviate from this line at higher values of $\rho_L \simeq 32\pi$ which may be due either to the beams becoming misaligned from the cold atoms as the AOM beam deflection increases with $\Delta f$, or may be an effect of the finite width of the pulses.  For $t_p=296~{\rm ns}$ the momentum boundary occurs at $\rho_b = 65\pi$, and as shown in \cite{Jones2004c} the maximum asymmetry due to the finite pulse width occurs at approximately $\rho_L = \frac{\rho_b}{2}$, and in the negative sense.  We should also note a dc offset to the data, which we believe arises from an initial misalignment of the laser beams, or from a systematic error in locating the centre of mass.  The signature of the kick-to-kick correlations, however, is the ac signal which is clearly shown in both sets fo data. 

To break spatial symmetry a linear `rocking' term of alternating sign is included by accelerating the optical lattice \cite{Madison1999}. This is done by modulating the frequency of one of the laser beams in a linear manner using a second (phase-locked) arbitrary function generator by an amount $\pm\delta\! f$ in the time of the kick period $T$. In the accelerating frame an inertial term appears in the Hamiltonian
\begin{equation}
H = \frac{p'^2}{2M} + V_0\cos(2k_L x') \pm max'
\label{AcceleratingHamiltonian}
\end{equation}
where the primes indicate variables in the accelerating frame. 
If, as before, this is now recast into dimensionless form we find that (dropping the primes for convenience)
\begin{equation}
\mathcal{H} = \frac{\rho^2}{2} + \bigg(K\cos(\phi) \pm A\phi\bigg)\sum\delta(\tau - n)
\end{equation}
where the dimensionless potential gradient is related to the magnitude of the frequency modulation (acceleration of the lattice) by $A = 2\pi t_p\delta\! f$ for finite pulses of width $t_p$. Accelerating the potential thus provides a simple way of controlling the magnitude of $A$ and hence controlling the phase shift of the momentum-dependent diffusion constant in order to make it locally asymmetric around zero momentum.

\begin{figure}
\includegraphics[width=1.0\columnwidth]{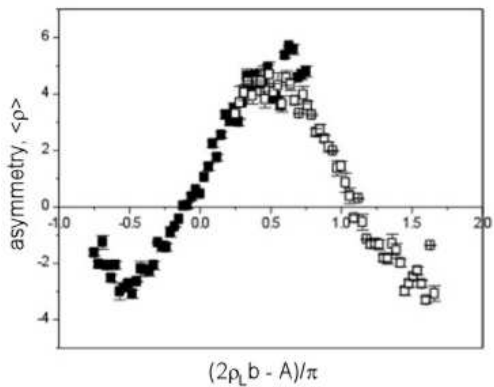}
\caption{Aysmmetry (first moment of momentum distribution) vs $(2\rho b-A)/\pi$\label{RockingRatchetGraph} for the accelerated lattice with chirped kicks experiment. Filled squares are data for $\rho_0 = 0$, open squares are $\rho_0 = 8\pi$. The asymmetry oscillates with a period of $2\pi$. The parameters for this experiment are $K$ = 2.6, $b$=1/16, $\hbar_{ef\!f}=1$ and 120 kicks.}
\end{figure}

For the accelerating lattice experiment the parameters were $K=2.6$, $T=9.47~\mu{\rm s}$, (so $\hbar_{ef\!f}=1$,) $t_p=296~{\rm ns}$ and $b=1/16$. The number of kicks was 120. As the maximum frequency modulation amplitude allowed by the radio-frequency synthesizers was $\pm 1.25~{\rm MHz}$ this limits the range of $A$ achievable to $\pm3\pi /4$. In order to observe one complete oscillation of the momentum diffusion constant, for some experiments an additional constant frequency offset was introduced between the laser beams such that in the rest frame of the lattice the mean atomic momentum was $\rho_L = 8\pi$. The asymmetry of the momentum distribution (calculated as above) was measured as a function of the amplitude of the frequency modulation of the laser beam, $\delta\! f$, for both $\rho_L = 0$ and $\rho_L = 8\pi$ and plotted as a function of $(2\rho_Lb-A)/\pi$. Results are shown in figure~\ref{RockingRatchetGraph} and can be seen to be sinusoidal (proportional to the local gradient of the diffusion constant) with a period of $2\pi$ as expected from the theory.

\begin{figure}
\includegraphics[width=1.0\columnwidth]{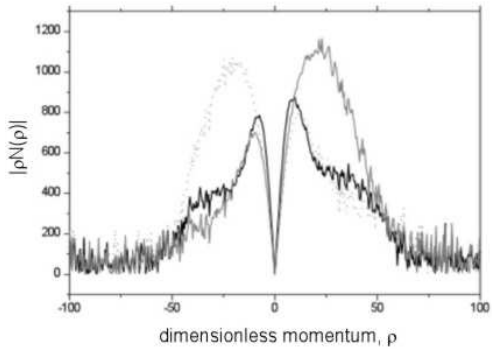}
\caption{Example momentum distributions from the accelerating lattice with chirped kicks experiment\label{RockingRatchetResults}. Following the example of Jonckheere \textit{et al.} we plot the modulus of the first moment of the momentum distribution, i.e. $\vert \rho  N(\rho)\vert$ for clarity. The black line is for $A=0$ and is almost symmetric. The solid grey line for $A=-\pi/2$ and the dotted grey line for $A=+\pi/2$, which develop large asymmetries in opposite senses.}
\end{figure}

Examples of the momentum distributions obtained from this experiment are shown in figure~\ref{RockingRatchetResults}. All three graphs are for $\rho_L =0$, and following the example of Jonckheere \textit{et al.} \cite{Jonckheere2003} we plot the modulus of the first moment of the momentum distribution $\vert\rho N(\rho)\vert$ for clarity. It can be seen that while the distribution for $A=0$ (black line in figure~\ref{RockingRatchetResults}) is almost symmetric $A=-\pi/2$ produces a large positive asymmetry and $A=+\pi/2$ a large negative asymmetry.

In conclusion we have shown that by breaking time and space symmetry in the delta-kicked rotor correlations between kicks give rise to a momentum-dependent diffusion rate.  We have exploited this in order to demonstrate experimentally a system that exhibits an asymmetric momentum diffusion due only to chaotic dynamics, in contrast to previous work that relies on specific features or structures in phase space.

\begin{acknowledgments}
We would like to thank past and present members of the UCL Quantum Chaos
Theory group for useful discussions and EPSRC and UCL for financial
support.
\end{acknowledgments}

\bibliography{JonesEtAl}

\end{document}